\documentclass[aps,prl,twocolumn,showpacs]{revtex4}

\usepackage{graphicx}
\usepackage{dcolumn}
\usepackage{bm}

\begin{document}


\title{Controlling Strong Electromagnetic Fields at a Sub-Wavelength Scale .}


\author{J. Le Perchec, P. Qu\'emerais$^{*}$, A. Barbara and  T. L\'{o}pez-R\'{\i}os}
\email[]{pascal.quemerais@grenoble.cnrs.fr}
\affiliation{Laboratoire d'Etudes des Propri\'et\'es Electroniques des
Solides,\\ (LEPES/CNRS), BP 166, 38042 Grenoble Cedex 9, France\\}


\date{\today}

\begin{abstract}
We investigate the optical response of two sub-wavelength grooves on
a metallic screen, separated by a sub-wavelength distance. We show
that the Fabry-Perot-like mode, already observed in one-dimensional
periodic gratings and known for a single slit, splits into two
resonances in our system : a symmetrical mode with a small Q-factor,
and an antisymmetric one which leads to a much stronger light
enhancement. This behavior results from the near-field coupling of
the grooves. Moreover, the use of a second incident wave allows to
control the localization of the photons in the groove of our choice,
depending on the phase difference between the two incident waves.
The system exactly acts as a \textit{sub-wavelength optical switch}
operated from far-field.
\end{abstract}

\pacs{71.36+c,73.20.Mf,78.66.Bz}

\maketitle Surface Enhancement Raman Scattering (SERS) still remains
a mystery in a large part, even though it is now accepted that the
excitation of localized electromagnetic modes of irregular metallic
surfaces is involved in its basic mechanism\cite{otto,moskovits}.
Optical excitation of such modes can indeed lead to important
concentration of electromagnetic energy in volumes (cavities) much
smaller than $\lambda^3$ where $\lambda$ is the excitation
wavelength, as it is the case for SERS active surfaces. These
specific places of very strong electromagnetic fields localization
are called "active sites" or "hot spots". However, the debate on the
origin of these hot spots remains open, as well as the hope to
control one day this phenomenon. The large interest raised by this
fundamental physics is also increased by its wide potential
applications in biochips, sensors, nano-antennae, optoelectronics or
energy transport on nanostructured surfaces.

In this letter, we consider a simple system which allows to
\textit{produce and control} the localization in space of such hot
spot phenomenon. It only consists of two deep grooves on a plane
metallic gold surface (fig.1). The excited modes appear, for the
chosen geometry, in the infrared region where we can consider the
metal as being a good reflector. Under this condition, a reliable
theoretical method, i.e the modal method using surface impedance
boundary conditions, can be used \cite{wirgin}. This method has
already demonstrated its ability to give a good qualitative and
quantitative agreement with the measured reflectivity of metallic
gratings \cite{barbara1,barbara2,barbara3}. The case of one groove
only was considered a long time ago \cite{deleuil}, while the
transmission for one \cite{takakura} and two slits \cite{schouten}
were only recently considered. In contrast with \cite{schouten}, the
distance between our two grooves is small with respect to the
incident wavelength. Very recently it also was shown \cite{skigin}
that sharp and deep resonances appear in the transmission response
of gratings with more than one slit per period or in gold dipole
antennas\cite{martin}. We here analyze the physical origin of this
new kind of resonances for a two slit system. As we will see, this
allows us to point out some very fundamental aspects of
electromagnetic resonances on metallic surfaces, and to control the
light localization by using a simple device.

We consider a p-polarized incident plane wave (electric field in the
plane of incidence) with a wavevector $k=2 \pi / \lambda$ impinging
on the surface at an angle $\theta$ (fig.1). The knowledge of the
magnetic field in the $z$-direction completely solves the problem as
$H_x=H_y=E_z=0$, $E_x=(i/ck\varepsilon_{0})\partial H_z/
\partial y$ and $E_y=(-i/ck\varepsilon_{0})\partial H_z/ \partial x$. In region (I), the field is expressed
as the sum of the incident wave and the reflected ones by:
\begin{displaymath}
\label{field} {H_{z}^{(I)}(x,y)} ={e^{ik(\sin \theta x-\cos \theta
y)}+\int^{+\infty}_{-\infty} {R(Q) e^{i(Qx+qy)} dQ}}
\end{displaymath}
where the distribution $R(Q)$ represents the amplitude of the
reflected field at the wavevector $(Q, q)$ with $q= \sqrt
{k^2-Q^2}$. In region (II) one has:
\begin{eqnarray}
{{H_{z}^{(II)}}(x,y)} &=& {A_1 [ e^{i k y}+ \alpha e^{-i k y} ]  I_1(x)}  \nonumber \\
&+& {A_2 [ e^{i k y}+ \alpha e^{-i k y} ]  I_2(x),}\nonumber
\end{eqnarray}
\begin{figure}
\centering\includegraphics[scale=0.4]{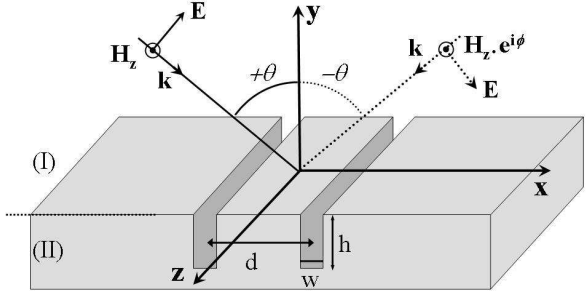}
\caption{Geometrical configuration and parameters. Region (I) and
(II) respectively corresponds to the region above and below the
metallic surface. $\phi$ is the phase difference between the two
incident waves described at the end of the paper.}
\end{figure}
where $I_1(x)$ (respectively $I_2(x)$) equal 1 in the interval
$[(-w-d)/2,(w-d)/2]$ (resp. $[(d-w)/2,(d+w)/2]$) and zero elsewhere.
$\alpha = [(1+ Z)/(1-Z)] \Phi^2$, with $\Phi =e^{-ikh}$, $Z = 1 /
\sqrt{\epsilon}$ is the surface impedance of the metal and
$\epsilon$ its dielectric constant. The expression for
$H_{z}^{(II)}$ assumes that the field is constant along $x$ within
each groove, which is a good approximation in the limit where $w<<
\lambda$ \cite{barbara1,barbara2}. To illustrate our results
numerically, we have fixed $w=0.2$ $\mu m$, $h=1.5$ $\mu m$, and
$d=0.5$ $\mu m$ all along the paper. The values of the complex
dielectric constant $\epsilon (\lambda)$ are taken from
\cite{handbook}. \\
The unknown variables are the distribution $R(Q)$ and the field
amplitudes $A_1$ and $A_2$ in the first and second groove
respectively. A set of equations is obtained by applying the
boundary conditions at the interface $y=0$ : $H_z^{(I)}=H_z^{(II)}$
at the mouth of each groove, and $\partial H_{z}^{(I)} /
\partial y +i kZ H_{z}^{(I)}=\partial H_{z}^{(II)} / \partial y +i k
Z H_{z}^{(II)}$ along the whole interface. After some elementary
algebra (see \cite{barbara2} for detailed procedure), the vector
${\bf{A} }= (A_1,A_2)$ is related to the excitation vector ${\bf
V}=(V_1,V_2)$ (null without the incoming wave), by the matricial
relation ${\bf A}={\bf M}^{-1} {\bf V}$, where ${\bf M}$ is the $2
\times 2$ symmetrical matrix which verifies $m_{11}=m_{22}$ with:
\begin{eqnarray}
{m_{11}}&=&{ \left ( 1+\alpha \right ) -  \Gamma (1+Z)(1-\Phi^2) \int_{-\infty}^{+ \infty} { \frac {\sec^2 \left(Q w/2 \right)}{q+kZ} } dQ}  \nonumber \\
{m_{12}}&=&{-\Gamma (1+Z)(1-\Phi^2) \int_{-\infty}^{+ \infty}
{\nonumber \frac {\sec^2 \left(Q w/2 \right)} {q+kZ} e^{iQd} dQ},}
\end{eqnarray}
where $\Gamma=w/\lambda$. The coordinates of the vector ${\bf V}$
are:
\begin{eqnarray}
{V_1}&=&{e^{-i \varphi} V_0} \quad\text{and} \quad \nonumber
{V_2}={e^{i\varphi}V_0} \quad \text{with:} \\ \nonumber
{V_0}&=&{\frac {2 \cos \theta} {\cos \theta+ Z} \sec \left(k \sin
(\theta) w/2 \right) \nonumber}
\end{eqnarray}
where we have introduced the angle $\varphi = k d \sin (\theta)/2$.
\\
The matrix $\bf M$ has two eigenvalues $m_{\pm}=-i(1-\Phi^2)e_{\pm}$
with respective eigenvectors ${\bf U}_{\pm} = (1, \pm 1)$ and:
\begin{eqnarray}
{e_{\pm}} &=& {\frac {1}{1-Z} \left( \cot (kh) - iZ \right ) - 2i \Gamma (1+Z) \times} \nonumber \\
                  & \times & { \int_{0}^{+ \infty}{(1 \pm \cos \left( k d u \right)) \frac {\sec^2(k w u/2)}{ \sqrt {1-u^2}+ Z}du}.\nonumber}
\end{eqnarray}
We have made the variable change $Q=ku$ in the integrals. The
solution of the problem is then:
\begin{eqnarray}
{A_{n=1,2}} &=& {\left [ \frac {1} {e_+}+(-1)^n i \tan \varphi \frac {1}{e_-} \right ] {\frac {i\cos(\varphi)}{1-\Phi^2}V_0}} \nonumber \\
{R(Q)} &=& {\frac {\cos \theta- Z} {\cos \theta+ Z} \delta (Q-k \sin \theta)+ \Gamma (1+Z)(1-\Phi^2) \times} \nonumber \\
            & \times & {(e^{iQd/2}A_1+e^{-iQd/2} A_2) \frac {\sec(Qw/2)}{q+kZ}}.
\end{eqnarray}
\begin{figure}
\centering\includegraphics[scale=0.4]{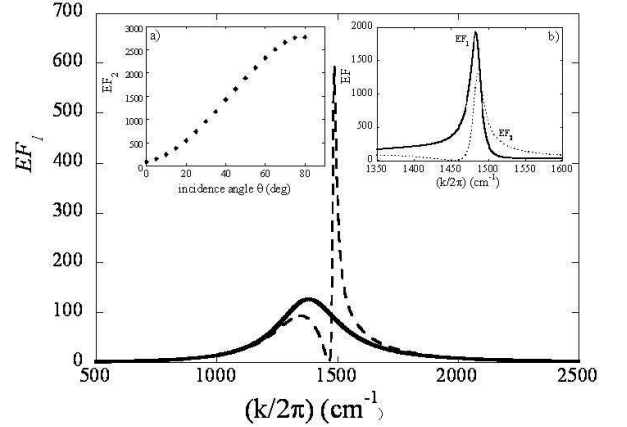} \caption{$EF_1$
of the cavity centered at $x=-d/2$ calculated for $\theta=0^\circ$
(full line) and $\theta=30^\circ$ (dotted line) as a function of the
wavenumber. Inset a) represents the behavior of $EF_2$ (cavity
centered at $x=+d/2$), as a function of the incidence angle $\theta$
and at 1484 $cm^{-1}$. Inset b) gives $EF_1$ and $EF_2$ of both
cavities calculated for $50^\circ$.}\label{fig-gratings}
\end{figure}
At the sight of eq.(1), one can see that the system presents two
electromagnetic resonances at $k=k_{\pm}$, which appear when $\Re
(e_+)=0$ and $\Re (e_-)=0$, with lineshapes respectively governed by
$\Im (e_+)$ and $\Im (e_-)$ ($\Re(x)$ and $\Im(x)$ being the real
and imaginary parts of $x$). The fields in the cavities are always a
linear combination of the two eigenvectors ${\bf A} \sim a_- {\bf
U}_-+a_+ { \bf U}_+$. However, when $k=k_+$ (respectively $k=k_-$),
the vector ${\bf A}$ is almost collinear with ${\bf U_+}$ (resp.
${\bf U_-}$) as the amplitudes in the two cavities are dominated by
the same (resp. the opposite) term. We will thus call the resonance
occurring at $k=k_-$ the ($-$) antisymmetric mode and that occurring
at $k=k_+$ the ($+$) symmetrical one. Contrary to the $(+)$ mode
which always exists, the $(-)$ one only develops for $\theta \neq 0$
(see fig.2) as it vanishes at normal incidence with tan$\varphi=0$.
Its bandwidth is much thinner than that of the symmetrical mode and
its enhancement factor is much larger. The enhancement factor
($EF$), defined as $|E_x/E_0|^{2}$ where $E_0$ is the incident
electric field, reflects the amount of stocked electromagnetic
energy at the resonances. For convenience, we note $EF_1$ and $EF_2$
the enhancement factors calculated at the mouth of each cavity, i.e
at $x=\pm d/2$ and $y=0$ where they are expressed as $EF_{n=1,2}=
|A_{n=1,2}(1-\alpha)|^2$. The $EF$ of the $(-)$ mode, shown in the
inset (a) of fig.2, increases with $\theta$ and its value can reach
more than $10^3$ whereas that of the symmetrical mode stays at
around $100$.
\\Another important point to highlight is that around the $(-)$
resonance, the fields in the two cavities are not strictly
identical. Inset (b) of fig.2 displays the $EF$ at the mouth of each
cavity close to $k=k_-$. At $1483$ $cm^{-1}$ the maximum of $EF_1$
is reached whereas the value of $EF_2$ is still low; at $1490$
$cm^{-1}$ $EF_1$ and $EF_2$ both take the same value. Around this
mode, the system thus develops a very high sensitivity: with a very
little variation of wavenumber (here less than one percent), the
field "jumps" from one cavity to the other. This behavior is
qualitatively comparable to the "unstable" behavior of hot spots
observed on SERS active surfaces.

In the following, we consider the metal as being a perfect
reflector, i.e $Z = 0$. This approximation induces only small
quantitative modifications and allows an analytical study which
highly helps to clarify the physics of the problem. however, the
presented numerical results are obtained without this approximation,
i.e using the finite value of $\epsilon (\lambda)$. We first compare
the two grooves system to the one where there is only one groove
centered at $x=0$. In this case, the amplitude of the field $A_0$ in
the \textit{unique cavity} is given by $A_0= i \left [ 1-\Phi^2
\right ]^{-1} V_0/e$, with :
\begin{displaymath}
e= \cot (kh) - 2 i \Gamma \int_{0}^{+\infty} {\frac {\sec^2 (k w
u/2)}{\sqrt {1-u^2}} du}.
\end{displaymath}
The resonance of this cavity occurs at $k=k_0= \omega_0/c$ for which
$\Re (e)=0$. Close to $k_0$, the field $A_0$ can be expanded around
$\omega_0$ as:
\begin{displaymath}
A_0 \approx \frac {C_0} {\omega_0-\omega - i \gamma_0/2},
\end{displaymath}
with $C_0 \approx i c V_0 /2h$, and where we have taken advantage of
the fact that at the resonance $k_0h \approx \pi/2$ \cite{barbara2}.
This equation is typical of a forced oscillator and, as the electric
field inside the cavity is proportional to $A_0$, indicates that the
cavity behaves as a forced oscillating dipole with a radiation
damping $\gamma_0 = 2w \omega_0^2/ \pi c$ and an effective
electromagnetic radius $r_0= 2w/\pi$. The effective dipolar
momentum, parallel to the interface, takes its maximum at the mouth
of the groove and decreases along the vertical walls. The maximum of
intensity at $\omega=\omega_0$ is $\vert A_0 \vert^2 \approx
4/(k_0w)^2$, typically of order 100 for our geometrical parameters.
We now expand, in the same manner, the values of $e_{\pm}$ around
the same $k_0$ for the two groove system. We easily get:
\begin{eqnarray}
{e_+} & \approx & {\left( \omega_+ - \omega - i \gamma_+ /2 \right) h/c} \\
{e_-} & \approx & {\left( \omega_- - \omega - i \gamma_- /2 \right)
h/c, \nonumber} \end{eqnarray} with ${\omega_{\pm}}={\omega_0 \mp
\Delta}$, ${\gamma_+}={2 \gamma_0}$ and ${\gamma_-}={\gamma_0 \left(
k_0 d/2 \right)^2}$. The shift $\Delta$, of the order of
$\gamma_0\ll\omega_0$, is:
\begin{equation}
 \Delta = \frac {\gamma_0} {\pi } \int_{1}^{+ \infty}{ \frac { \cos \left( k_0 d u \right) \sec^2 \left( k_0 w u/2 \right)} {\sqrt {u^2-1}} du}
\end{equation}

\begin{figure}
\centering\includegraphics[scale=0.4]{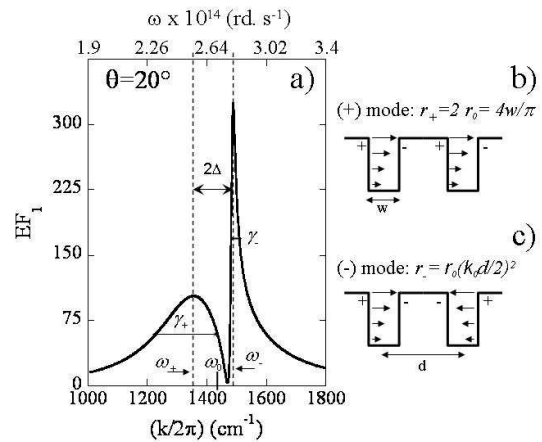} \caption{$EF_1$
calculated for $\theta=20^\circ$ showing the resonances $(+)$ and
$(-)$ characterized by their eigenfrequencies $\omega_\pm$, located
on both sides of the frequency resonance $\omega_0$ of a unique
cavity, and their bandwidth $\gamma_\pm$ (a). Right part
schematically represents the in-phase coupling of the $(+)$ mode (b)
and the anti-phase coupling of the $(-)$ mode (c) and their
corresponding equivalent dipole.}
\end{figure}
Eq. (2, 3) confirm our numerical observation as they show that the
width of the $(-)$ mode, driven by $\gamma_-$, is much lower than
that of the $(+)$ mode, driven by $\gamma_+$, owing to the small
factor $(k_0d)^2$ (and recalling our sub-wavelength coupling
hypothesis : $\lambda_0>> d$). A physical image of these resonances
can be given noticing that our results are completely similar to
those obtained by Lyuboshitz\cite{lyuboshitz} for \textit{two
near-field coupled oscillating dipoles}. Our resonances thus arise
from the near-field coupling of two identical grooves, individually
resonating at $\omega_0$. The symmetrical $(+)$ mode corresponds to
the in-phase oscillation of each cavity whereas the second one
corresponds to an anti-phase oscillation. The distribution of
electric field in the cavities for each mode is sketched in fig.3.
As a consequence of this coupling, the $(+)$ mode has a strong
dipolar character with an effective dipolar moment close to twice
that of a unique cavity and a large electromagnetic radius $r_+
=2r_0$. On the opposite, the $(-)$ mode has an effective dipolar
moment almost null,  with a much smaller electromagnetic radius
$r_-=r_0 (k_0d/2)^2$, and its radiation pattern is essentially that
of a quadripole. This explains why this mode is weakly radiative and
with an extremely narrow lineshape, very different from the
width of the in-phase mode. \\
Searching for the location of the maximum of the field in each
cavity around the $(-)$ mode, ones gets for non zero $\theta$:
\begin{eqnarray}
\omega_{max.}^{n=1,2} &\approx& \omega_- - \frac {(-1)^{n} \left(
\frac {k_0d}{2} \right)^3 \gamma_0} {4\sin
\theta\left(1+\left(\frac{2\Delta}{\gamma_0}\right)^{2}\right)}
 + O((k_0d)^4) \nonumber \\
\vert A_{max} \vert^2 &\approx& \frac {16 \sin^2 \theta} {(k_0w)^2
(k_0d)^2}, \nonumber
\end{eqnarray} where $\vert A_{max} \vert^2$ is proportional to the intensity of the
field in both cavities at $\omega_{max}$. The two maxima
$\omega_{max}^1$ and $\omega_{max}^2$ are separated by a very small
frequency difference of the order of $(k_0d)^3 \gamma_0$, which,
together with the narrow lineshape of the resonance, explains why
the profile of the field strongly varies in this region. The
magnitude of $\vert A_{max} \vert^2$ requires some comment. Indeed,
for an usual oscillator with damping $\gamma$, the maximum of
intensity of the oscillation scales as $\gamma^{-2}$, so that $\vert
A_{max} \vert ^2$ should scale as $\gamma_-^{-2} \sim (k_0d)^{-4}$
instead of $(k_0d)^{-2}$. The field intensity of the $(+)$ mode
scales, as expected, as $\gamma_+^{-2}$ (eq. 1). The reason for that
is that the ($+$) and $(-)$ modes {\it are not sensitive to the same
parts of the incident electric field}. Since $d/\lambda \ll 1$, the
latter can express at interface as $E_0(1+ikx)$ at the scale of our
two-grooves system. The even term corresponding to the mean value of
the field excites the $(+)$ mode and the odd one, corresponding to
the local variations of the field, excites the $(-)$ mode. This mode
is thus sensitive to an "effective" field of intensity $\sim
E_0^2(k_0d)^2$ at the mouth of the grooves, whereas the $(+)$ mode
is excited by an effective field of intensity $E_0^2$. This is the
origin of the lost of a factor $(k_0d)^{-2}$ in the intensities of
the mode $(-)$. The latter results from a strong resonator, but
excited by a very weak effective field.

We now take advantage of our understanding to control - from far
field - the light localization in the cavity of our choice, or in
both. To do so, we introduce a new free parameter by sending a
second incident plane wave, at the same frequency, with an incidence
angle $-\theta$, and with a phase difference $\phi$ with respect to
the first incident wave (fig.1). Changing $\phi$, we can control the
incident effective fields respectively exciting one mode or the
other. Different states, that we code as : $(1,1)$, $(1,-1)$,
$(1,0)$ and $(0,-1)$ can be achieved. The first two, $(1,1)$ and
$(1,-1)$ respectively correspond to the case where only the pure
$(+)$ or only the pure $(-)$ resonances are excited. The cavities
are then completely in-phase or in anti-phase. The two other ones
correspond to cases where \textit{one of the cavities} is lit
(cavity 1 for $(1,0)$, and cavity 2 for $(0,-1)$). As $\phi$ is a
parameter easy to modify, for instance changing the optical path, we
can control in straightforward manner the field localization.
\\With two incoming waves, the field becomes:
\begin{displaymath}
H_{inc.}=\left[e^{ik \sin \theta x}+e^{i \left( \phi-k \sin \theta x
\right)}\right] e^{-ik \cos \theta y},
\end{displaymath}
and the solution for each cavity can be written as:
\begin{equation}
A_{n=1,2} \sim \frac {\cos (\phi/2)} {e_+}+(-1)^n  \tan \varphi  \frac {\sin (\phi/2)}{e_-} \\
\end{equation}
where we did not write explicitly some unimportant prefactor common
to both cavities. From these equations, it is easy to see that for
$\phi=\phi_{(1,1)}=0$, one gets $A_1=A_2 \sim 1/e_+$, so that at
$k=k_+$ we have the pure $(+)$ resonance. In the same manner, the
pure $(-)$ resonance can be excited at $k=k_-$ when $\phi =
\phi_{(1,-1)}= \pi$ where $A_1=-A_2 \sim - \tan \varphi/e_-$. This
last state presents an extremely high $EF \sim 10^4$ at
$\theta=80^{\circ}$.
\\More subtle is the possibility to control the
extinction of the field in only one of the cavities (of our choice)
while the other one is resonating.
\begin{figure}
\centering\includegraphics[scale=0.4]{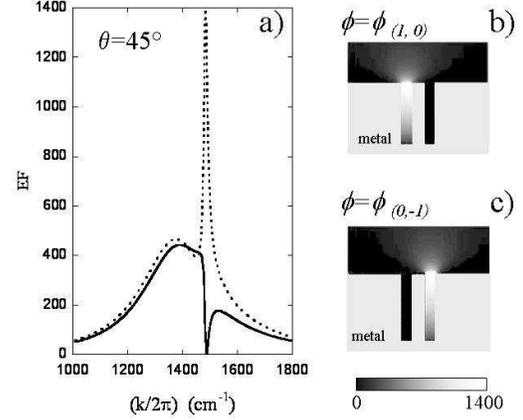} \caption{$EF_1$
and $EF_2$ spectra at $\theta=45^\circ$ for two incoming waves (a)
and related mappings of the electric field amplitude $E_x$ (b, c) at
1490 $cm^{-1}$. For $\phi=\phi_{(1,0)}$ the first cavity is lit,
$EF_1$ is the dotted line, and the second cavity is extinguished,
$EF_2$ is the full line. For $\phi=\phi_{(0,-1)}$ it is the
opposite.} \label{fig-gratings}
\end{figure}
Eq.(4) shows that this can be achieved provided that $\cot
\phi/2=\pm (e_+/e_-) \tan \varphi$, the sign $"+"$ corresponding to
the $(0,-1)$ state and the $"-"$ sign to the $(1,0)$ state. This
condition can be satisfied (since the function $\cot (x)$ can vary
from $-\infty$ to $+\infty$), provided that $e_+/e_-$ is real. This
is obtained for $\omega \approx \omega_- + 2 \Delta
(\gamma_-/\gamma_+)$, which is very close to $\omega_-$. Figure 4
represents the $EF$ of both cavities either choosing
$\phi=\phi_{(1,0)}$ or $\phi=\phi_{(0,-1)}$, together with the
related mappings of the electric field amplitude $E_x$. These show
how the field can be strongly localized in only one of the cavity,
while the second one is completely extinguished and this, even
though the cavities are identical and separated by a sub-wavelength
distance.\\In conclusion, we have demonstrated that the near-field
coupling of two metallic resonating cavities leads to a resonance
with an extremely thin spectral width, which can localize very
intense fields. This could be a key point in the understanding of
the SERS, as the described physics should remain valid in the
visible region, except for a scaling factor. Finally, we proposed a
simple way to control the near-field of each cavity, enabling this
system to act as a sub-wavelength optical switch simply operated
from the far-field.

\begin{acknowledgments}
This work is partly the result of illuminating discussions of P.
Qu\'emerais with D. Mayou which have been very beneficial to us. We
also would like to thank P. Lalanne for helpful conversations about
electromagnetic resonances in gratings.
\end{acknowledgments}


\end{document}